\documentclass{aa}    %% Astronomy & Astrophysics style class aa.cls v8.2
\usepackage{graphicx,twoopt,natbib}
\usepackage[varg]{txfonts}           %% old-fashioned A&A font choice
\usepackage{hyperref}                %% for pdflatex
\hypersetup{
  colorlinks=true,   %% links colored instead of frames
  urlcolor=blue,     %% external hyperlinks
  linkcolor=blue     %% internal latex links (eg Fig)
}

\bibpunct{(}{)}{;}{a}{}{,}    %% natbib cite format used by A&A and ApJ

\graphicspath{ {figures/} }

\makeatletter
\newcommand{\bibnote}[2]{\global\@namedef{#1note}{#2}}
\newcommand{\biblink}[2]{\global\@namedef{#1link}{#2}}
\newcommand{\Alfven}[1]{Alfv\'en}

\makeatother

\makeatletter
  \protected\def\stonyslink{%
     \def\hyper@linkstart##1##2{}\let\hyper@linkend\@empty}
  \newcommandtwoopt{\citeads}[3][][]{%
   \href{http://ui.adsabs.harvard.edu/abs/#3/abstract}%
        {\stonyslink \citealp[#1][#2]{#3}}%   %% Rutten, 2000
   \biblink{#3}{\href{http://ui.adsabs.harvard.edu/abs/#3/abstract}{ADS}}}
 \newcommandtwoopt{\citepads}[3][][]{%
   \href{http://ui.adsabs.harvard.edu/abs/#3/abstract}%
        {\stonyslink \citep[#1][#2]{#3}}%     %% (Rutten 2000)
   \biblink{#3}{\href{http://ui.adsabs.harvard.edu/abs/#3/abstract}{ADS}}}
 \newcommandtwoopt{\citetads}[3][][]{%
   \href{http://ui.adsabs.harvard.edu/abs/#3/abstract}%
        {\stonyslink \citet[#1][#2]{#3}}%     %% Rutten (2000)
  \biblink{#3}{\href{http://ui.adsabs.harvard.edu/abs/#3/abstract}{ADS}}}
 \newcommandtwoopt{\citeyearads}[3][][]{%
   \href{http://ui.adsabs.harvard.edu/abs/#3/abstract}%
        {\stonyslink \citeyear[#1][#2]{#3}}%  %% 2000
   \biblink{#3}{\href{http://ui.adsabs.harvard.edu/abs/#3/abstract}{ADS}}}
\makeatother

\begin{document}  

%% Simple header (replacing A&A commands which produce the A&A banner)

   \title{Flux-tube-dependent propagation of \Alfven{A} waves in the solar corona}

   %\subtitle{Subtitle}

 \author{Chaitanya Prasad Sishtla \and
          Jens Pomoell \and
          Emilia Kilpua \and
          Simon Good \and
          Farhad Daei \and
          Minna Palmroth
          }
%   \author{Chaitanya Prasad Sishtla\inst{1} \and
%           Jens Pomoell\inst{2} \and
%           Emilia Kilpua\inst{3} \and
%           Simon Good\inst{4} \and
%           Farhad Daei\inst{5} \and
%           Minna Palmroth\inst{6}
%           }

  \institute{Department of Physics, University of Helsinki,
              Helsinki, Finland\\
              \email{chaitanya.sishtla@helsinki.fi}
  }
%   \institute{Department of Physics, University of Helsinki,
%               Helsinki, Finland\\
%               \email{chaitanya.sishtla@helsinki.fi}
%         \and
%              Department of Physics, University of Helsinki,
%               Helsinki, Finland\\
%               \email{jens.pomoell@helsinki.fi}
%         \and
%              Department of Physics, University of Helsinki,
%               Helsinki, Finland\\
%               \email{emilia.kilpua@helsinki.fi}
%         \and
%              Department of Physics, University of Helsinki,
%               Helsinki, Finland\\
%               \email{simon.good@helsinki.fi}
%         \and
%              Department of Physics, University of Helsinki,
%               Helsinki, Finland\\
%               \email{farhad.daei@helsinki.fi}
%         \and
%              Department of Physics, University of Helsinki,
%               Helsinki, Finland\\
%               \email{minna.palmroth@helsinki.fi}
%              }

   \date{}

% \abstract{}{}{}{}{} 
% 5 {} token are mandatory
 
  \abstract
   {\Alfven{a}-wave turbulence has emerged as an important heating mechanism to accelerate the solar wind. The generation of this turbulent heating is dependent on the presence and subsequent interaction of counter-propagating \Alfven{a} waves. This requires us to understand the propagation and evolution of \Alfven{a} waves in the solar wind in order to develop an understanding of the relationship between turbulent heating and solar-wind parameters.}
   {We aim to study the response of the solar wind upon injecting monochromatic single-frequency \Alfven{a} waves at the base of the corona for various magnetic flux-tube geometries.}
   {We used an ideal magnetohydrodynamic (MHD) model using an adiabatic equation of state. An \Alfven{A} pump wave was injected into the quiet solar wind by perturbing the transverse magnetic field and velocity components.}
   {\Alfven{A} waves were found to be reflected due to the development of the parametric decay instability (PDI). Further investigation revealed that the PDI was suppressed both by efficient reflections at low frequencies as well as magnetic flux-tube geometries.}
   {}
   

   \maketitle
%
%-------------------------------------------------------------------

%%%%%%%%%%%%%%%%%%%%%%%%%%%%%%%%%%%%%%%%%%%%%%%%%%%%%%%%%%%%%%%%%%%%%%%%%%%%
\section{Introduction}     \label{sec:introduction}
%%%%%%%%%%%%%%%%%%%%%%%%%%%%%%%%%%%%%%%%%%%%%%%%%%%%%%%%%%%%%%%%%%%%%%%%%%%%
The first studies on the formation and acceleration of the solar wind were performed by E. N. Parker in the late 1950s~\citep{parker1958dynamics, parker1960hydrodynamic}. Subsequently to the seminal work, considerable effort has been devoted to understanding the physical processes involved in the heating  of the solar wind, such as the energy partitioning between electrons and ions%incorporated separate energy equations for electrons and ions
~\citep{hartle1968two}, collisionless contributions to the electron heat flux~\citep{hollweg1974electron}, the role of the super-radial expansion of solar wind magnetic flux tubes~\citep{kopp1976dynamics, holzer1980conductive}, and energy transfer to the solar wind through shock (compressional) heating and \Alfven{a}-wave turbulence~\citep{coleman1968turbulence, alazraki1971solar, belcher1971large}.
Despite decades of research, fundamental questions regarding the processes generating the solar wind remain unresolved~\citep{gombosi2018extended, bruno2013solar, chandran2018parametric}. The strong magnetic fields carried by the solar wind interact with photospheric convective motions to generate \Alfven{a} waves~\citep{cranmer2005generation}. These \Alfven{a} waves are ubiquitous in in situ solar wind observations in interplanetary space \citep[e.g.][]{Belcher1971,dAmicis2015}. In the corona, their presence  has been confirmed by non-thermal line width~\citep{banerjee2009signatures, hahn2013observational} and Faraday-rotation fluctuations~\citep{hollweg1982possible, hollweg2010coronal}.

%The physical processes concerned with the heating and acceleration of the solar wind continues to garner much attention from the solar physics community~\citep{chandran2010alfven, reville2018parametric, shoda2019three}. Both are also (at least partly) still unresolved major research questions. 

The expanding solar wind also exhibits a turbulent character that in some respects resembles the well known hydrodynamic turbulence described by \citet{kolmogorov1991local}. In particular, a key element in the \Alfven{a}ic turbulence is the presence of counter-propagating \Alfven{a} waves, which non-linearly interact to form the turbulent cascade~\citep{goldreich1995toward}. \Alfven{a} waves generated due to photospheric motions propagate away from the Sun, but there are several suggested mechanisms on how to generate sunward travelling waves. First, the inhomogeneity of the solar corona causes \Alfven{a} waves to partially reflect, which enables turbulent heating~\citep{Ferraro1958, Shoda2016, an1990reflection, velli1993propagation, cranmer2005generation, verdini2007alfven}. Recent studies~\citep{magyar2017generalized} have shown that phase mixing caused by the \Alfven{a} velocity being inhomogenous perpendicular to the magnetic field lines~\citep{heyvaerts1983coronal, de2002fast, goossens2012surface} can also generate turbulent heating. %Such counter-propagating waves can  arise when forward travelling Alfven{a} waves that are generated as a response of convective plasma motion at the photosphere reflect in the corona due to gradual variations in the \Alfven{a}n speed  \citep[e.g.,][]{Ferraro1958,Shoda2016}.
% nonuniform media
Finally, large amplitude \Alfven{a} waves are subject to the parametric decay instability (PDI) in compressible and low beta plasma, resulting in \Alfven{a}-wave turbulence~\citep{gary2001plasma, iwai2014coronal, fu2018parametric, hoshino1989time, goldstein1978instability, del2015parametric}. 

The theory of PDI was already established in the 1970s \citep[e.g.][]{sagdeev1969nonlinear, Derby1978, goldstein1978instability}, but its importance in relation to solar-wind generation has continued even in recent studies~\citep[e.g.][]{shoda2018frequency, chandran2018parametric}. The PDI results in forwards (anti-sunward) propagating \Alfven{a} waves decaying into backwards (sunward) propagating \Alfven{a} waves and compressive forwards propagating ion acoustic waves. The forwards-propagating \Alfven{a} wave is often called a `pump wave'. The instability thus has a two-fold importance for solar-wind heating. First, backwards-propagating \Alfven{a} waves can lead to a $1/f$ turbulent cascade as they interact with the forwards propagating \Alfven{a} waves; secondly, ion acoustic waves can undergo Landau damping, effectively dissipating energy in MHD scales~\citep{fu2018parametric}. There is observational evidence for the presence of PDI both in the solar wind~\citep{bowen2018density} and in laboratory conditions~\citep{dorfman2016observation}. %This process also accounts for inverse cascade of energy, resulting in longer-period (low-frequency) \Alfven{a} waves from low-period (high-frequency) ones. This is also a critical element as longer-period \Alfven{a} waves are efficiently reflected in the transition region, while they dominate the solar wind spectrum \citep[e.g.,][]{leroy1981,reville2018parametric}. The PDI  has been recently observed both in the solar wind~\citep{bowen2018density} and in  laboratory conditions~\citep{dorfman2016observation}. The instability has also been found to occur dominantly in low-beta plasma, a condition generally met in particular in the low corona~\citep{fu2018parametric, hoshino1989time, goldstein1978instability}. 

The importance of the PDI in heating the solar wind has been studied through numerical simulations, which have demonstrated that solar-wind acceleration and coronal heating can be explained self-consistently via the PDI~\citep{matsumoto2014connecting, shoda2019three}. Subsequently, \cite{shoda2018self} also explored the relative dominance of turbulent and compressive heating in the solar corona. The salient feature of the numerical model employed in such studies is the injection of \Alfven{a} waves (i.e. pump waves) into the system through boundary conditions, which enables investigations pertaining to the linear and non-linear dynamics of these waves as they propagate and self-consistently interact to accelerate the plasma, thus forming the solar wind~\citep{shoda2018high, shoda2018frequency}. These studies have revealed that pump waves at certain frequencies, which are closely connected to the photospheric convective motions~\citep{cranmer2005generation}, significantly affect the occurrence of the PDI and subsequent \Alfven{a}-wave turbulence. It is important to note that such studies are fundamentally 
%The numerical formulation employed in such 
%investigations 
in contrast to other (global) theoretical studies 
that aim to provide an accurate model of the solar corona and solar wind by incorporating 
phenomenological prescriptions of \Alfven{a}-wave reflection-driven turbulent heating through extensions of the  Wentzel-Kramers-Brillouin (WKB) approximation~\citep{dewar1970interaction, van2014alfven, chandran2010alfven}.

%In this paper, we discuss the propagation of \Alfven{a} waves in the solar corona in the context of different flux tube expansion factors. 
An important parameter of interest in simulation studies of the solar wind is the rate of magnetic flux-tube expansion due to its correlation with terminal solar wind speeds~\citep{withbroe1988temperature, wang1990solar}. This rate of magnetic flux tube expansion can be derived through potential field extrapolations using magnetogram data, which evaluate this expansion factor at the fixed source surface height. However, more recent works~\citep{suzuki2006forecasting, pinto2017multiple} suggest that the terminal wind speeds can be better explained using both magnetic field amplitudes at the foot-point of the flux tubes and their expansion factors. The following question therefore remains: how do the dynamics of \Alfven{a} waves depend on the magnetic flux-tube expansion factor?

%The correlation between solar wind speeds observed at 1~AU and the rate of magnetic flux tube expansion in the corona is known to since the initial observation ~\citep{withbroe1988temperature, wang1990solar}. As a consequence of these findings, the rate of magnetic flux tube expansion is an important parameter of interest in simulation studies of the solar wind, particularly with the continued usage of potential field extrapolations using magnetogram data which evaluate this expansion factor at the fixed source surface height. However, more recent works~\citep{suzuki2006forecasting, pinto2017multiple} suggest that the terminal wind speeds can be better explained using both magnetic field amplitudes at the foot-point of the flux tubes and their expansion factors.  Additionally, the Parker Solar Probe \citep[PSP][]{Fox2016} affords us an unprecedentedly close view of the Sun, allowing us to directly investigate the importance of PDI for solar wind dynamics. 

In this study we investigated the behaviour of density perturbations and Els\"{a}sser variables to provide an understanding of the onset of PDI in the solar wind for different flux-tube expansion factors and pump frequencies. Although previous works~\citep{reville2018parametric, shoda2018frequency} have studied the response of the solar wind for various pump frequencies, only limited studies have been performed to observe the dependency of \Alfven{a} wave propagation on the expansion factors. This forms an important line of enquiry, as the expansion factors affect the non-linear evolution of \Alfven{a} waves as they propagate in the solar wind~\citep{hoshino1989time, farahani2011nonlinear, farahani2012nonlinear}.

%%%%%%%%%%%%%%%%%%%%%%%%%%%%%%%%%%%%%%%%%%%%%%%%%%%%%%%%%%%%%%%%%%%%%%%%%%%%
\section{Methodology}    \label{sec:methodology}
%%%%%%%%%%%%%%%%%%%%%%%%%%%%%%%%%%%%%%%%%%%%%%%%%%%%%%%%%%%%%%%%%%%%%%%%%%%
To model the response of the solar corona to perturbations originating in the photosphere, we solved for the solar wind dynamics in a single flux tube starting from the low corona and extending to $30$ solar radii ($R_\odot$) from the Sun. We considered a one-dimensional system with the geometry of the flux tube determined by its cross-sectional area $a(r)$, which varies with the distance along the flux tube $r$.

\subsection{Flux-tube geometry}

Similarly to the seminal work of~\citet{kopp1976dynamics}, we parametrise
the flux-tube geometry by specifying the cross-sectional area ($a$) 
to be proportional to the flux tube expansion $f$,
\begin{align}
 a = a_0\left(\frac{r}{r_0}\right)^2 f,
\end{align}
where $a_0$ is the cross-sectional area at the reference height $r_0$. By flux conservation, the magnetic field $\mathbf{B}$ satisfies
\begin{align}
    B_r = B_r(r_0) \frac{a_0}{a}.
\end{align}
Thus, $f$ describes the deviation of the expansion of the flux tube from that
of a radially expanding flux tube for which $a \propto r^2$. The expansion factor defined by~\citep{kopp1976dynamics} is used:
\begin{align}
    f = \frac{f_\mathrm{max}\exp\left(\left(r-R_1\right)/\sigma_1\right) + f_1}{\exp\left(\left(r-R_1\right)/\sigma_1\right) + 1},
    \label{eq:f}
\end{align}
with $f_1$ a constant chosen so that 
$f(r_0) = 1$.
%given by
%
%\begin{align}
%    f_1 = 1-(f_\text{max}-1)\exp((R_\odot - R_1)/\sigma_1).
%\end{align}
%
This parametrisation describes a flux tube where $f$ increases from 1 at $r=r_0$ and saturates to $f_\mathrm{max}$ at $r>>R_1$ with most variation occurring between $R_1 - \sigma_1$ to $R_1 + \sigma_1$. Here we take $R_1 = 1.3~R_\odot$, $\sigma_1 = 0.5~R_\odot$, and $r_0 = 1.014~R_\odot$.

To study the dependence of the dynamics for different flux tube geometries, we vary $f_\mathrm{max}$ and consider three different scenarios 
corresponding to $f_\mathrm{max} = 3, 5,~$and $10$. Such a flux-tube geometry has previously been used in other works to study solar wind properties in various numerical schemes~\citep{shoda2018frequency, shoda2018self, suzuki2005making, chandran2010alfven}. 
% \ToDo{Add note that cites some of the other work that uses the same geometry. This is at least shoda (which paper?), but also Chandran. Any others?}

\subsection{Governing equations}

The dynamical evolution of the plasma is modelled considering an ideal magnetohydrodynamic (MHD) description augmented with additional relevant physical processes such as gravity and an ad hoc energy source. The equations solved are the following:
\begin{align}
    \frac{\partial}{\partial t}\rho + \frac{1}{a}\frac{\partial}{\partial r}\left(a\rho v_r\right) = 0,
\label{eg:density}
\end{align}
\begin{multline}
    \frac{\partial}{\partial t}\left(\rho v_r\right) + \frac{1}{a}\frac{\partial}{\partial r}\left[a\left(\rho v_r^2 + p + \frac{\mathbf{B^2_\perp}}{2\mu_0}\right)\right] \\= \left(p + \frac{\rho \mathbf{v^2_\perp}}{2}\right)\frac{1}{a}\frac{\partial}{\partial r} a - \rho g,
\end{multline}
\begin{multline}
%\begin{split}
    \frac{\partial}{\partial t}\left(\rho \mathbf{v_\perp}\right) + \frac{1}{a}\frac{\partial}{\partial r}\left[a\left(\rho v_r\mathbf{v_\perp} - \frac{B_r\mathbf{B_\perp}}{\mu_0}\right)\right] \\= - \frac{1}{2a}\frac{\partial a}{\partial r}\left(\rho v_r\mathbf{v_\perp} - \frac{B_r\mathbf{B_\perp}}{\mu_0}\right),
%\end{split}
\end{multline}
\begin{align}
    \frac{1}{a}\frac{\partial}{\partial r}\left(aB_r\right) = 0,
\end{align}
\begin{align}
\begin{split}
    \frac{\partial}{\partial t} \mathbf{B_\perp} + \frac{1}{a}\frac{\partial}{\partial r}\left[a(\mathbf{B_\perp}v_r - B_r\mathbf{v_\perp}) \right] &= \frac{1}{2a}\frac{\partial a}{\partial r}\left(\mathbf{B_\perp}v_r - B_r\mathbf{v_\perp}\right),
\end{split}
\end{align}
%\begin{align}
%    \frac{1}{a}\frac{\partial}{\partial r}(aB_r) = 0
%\end{align}
\begin{multline}
%\begin{split}
    \frac{\partial}{\partial t}e 
    + \frac{1}{a}\frac{\partial}{\partial r}
      \left[a \left(v_r \left\{ e + p + \frac{\mathbf{B^2}}{2\mu_0}- \frac{B_r^2}{\mu_0} \right\} 
                   - B_r\frac{\mathbf{B_\perp \cdot v_\perp}}{\mu_0} \right)\right]
    \\= -\rho gv_r + S,
%\end{split}
\label{eg:energy}
\end{multline}
%\begin{align}
%    \text{where }~e = \frac{p}{\gamma-1} + \frac{\rho\mathbf{v^2}}{2} + %\frac{\mathbf{B^2}}{2\mu_0} 
%\end{align}
%
corresponding to the mass continuity equation, momentum, and induction equations for components parallel and perpendicular to the flux tube, and the energy equation, respectively.
The quantities $\rho$, $\mathbf{v}$, $\mathbf{B}$, $e$, and $p$ correspond to the mass density, plasma bulk velocity, magnetic field, total energy density, and thermal pressure, respectively. The directions along and transverse to the flux tube 
%radial and transverse directions 
are denoted by $r$ and $\perp$, respectively. The total energy density $e$ is composed of the thermal, kinetic, and magnetic energy densities:
\begin{align}
    e = \frac{p}{\gamma-1} + \frac{\rho\mathbf{v}^2}{2} + \frac{\mathbf{B}^2}{2\mu_0}.
\end{align}
A polytropic index of $\gamma = 5/3$ appropriate for a monoatomic gas is assumed. We also include gravity with the gravitational acceleration $g=\frac{G M_\odot}{R_\odot^2}$. In order to accurately simulate a solar wind profile for an adiabatic $\gamma,$ we incorporate an additional coronal energy source term~$S$~\citep{pomoell2015modelling, mikic2018predicting} to obtain 
a steady-state solar wind that approximates a Parker-like outflow. To that end, we specify a static energy source term:
%a realistic heating 
%and compression of the solar wind plasma,
%of the solar wind
%
\begin{align}
    S = S_0 \mathrm{exp}\left(-\frac{r}{L}\right),
\end{align}
with $S_0 = 0.5\times 10^{-6}$~Wm$^{-3}$ and $L = 0.4R_\odot$~m. 
We note that as a simplification, we omit thermal conduction from our model, which has an important role in the redistribution of energy in the low corona. Instead, we set the base of our modelling domain to be located in the corona, which is characterised by a high temperature and low density.
%Another important physical process that is relevant for the heating of the solar wind is thermal conduction~\citep{matsumoto2014connecting} which we omit from our numerical model. It is responsible for transferring heat outwards, and skewing the ion and electron distribution functions causing the generation of ion-acoustic, magneto-acoustic, and ion-cyclotron waves~\citep{forslund1970instabilities}. Thus, while it is important to incorporate heat conduction to obtain realistic solar wind processes in our simulation, it has limited effect on the qualitative comparisons on the propagation of pump waves in different flux tube geometries made in our study.
%\ToDo{Discuss this more at length here or in 2.1?}

\begin{figure}[ht]
\centering
\includegraphics[width=0.5\textwidth]{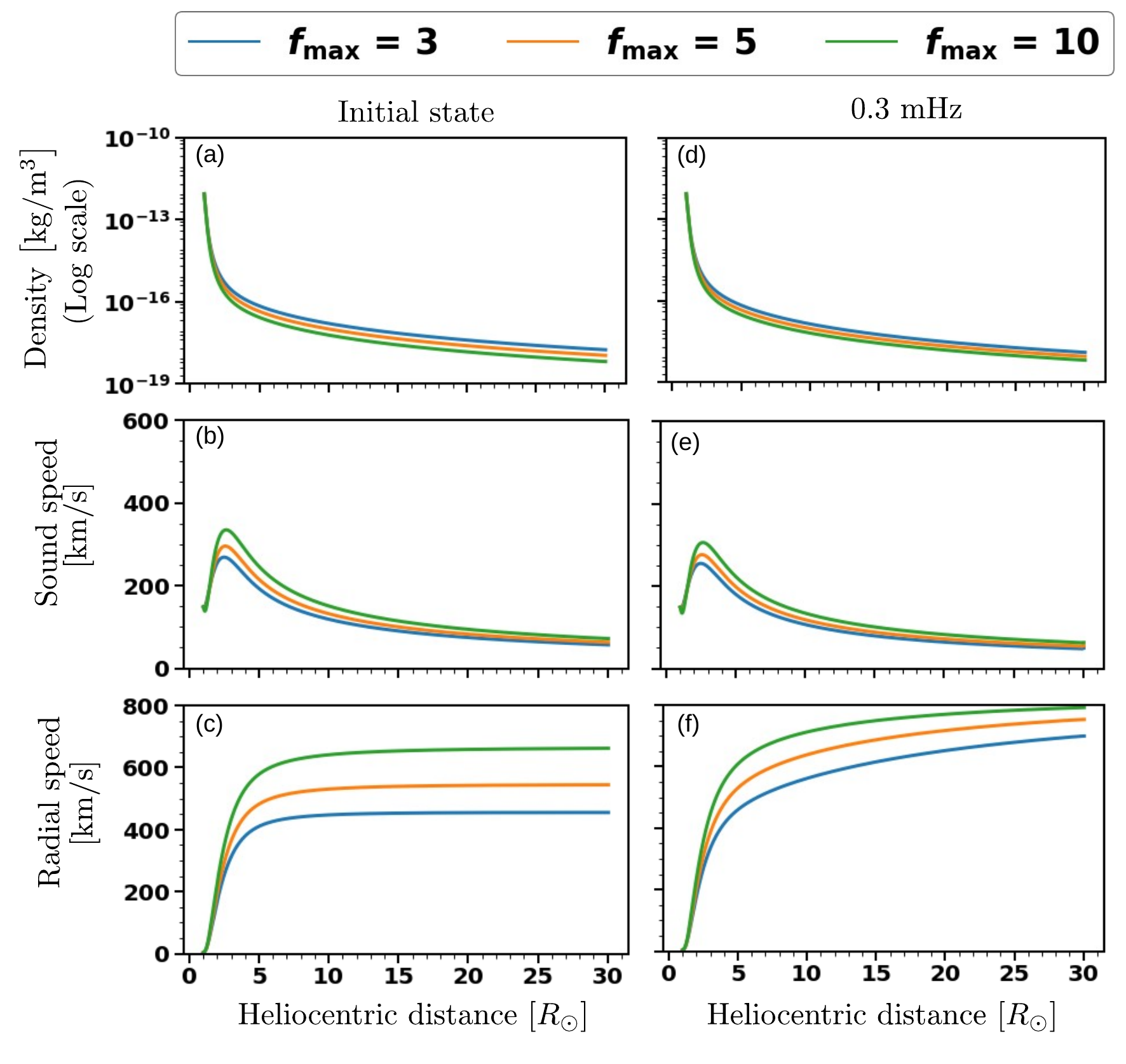}
\caption{Panels a-c present the mass density, sound speed, and radial speed of the solar wind before any injection of a pump wave. The solar-wind response after the injection of a 0.3~mHz pump wave is similarly shown in Panels d-f. The solar-wind solutions are presented for the different magnetic flux tube expansion factors: $f_\mathrm{max} = 3, 5, 10$.}
\label{fig:init-cond}
\end{figure}

%\subsection{Steady-state solar wind}    \label{subsec:methodology-BC}
\subsection{Simulation domain, boundary conditions, and computation stages}    \label{subsec:methodology-BC}

The governing equations are integrated forward in time in a spatial domain spanning from the low corona to the solar wind. Specifically, we choose the heliocentric distance of the inner boundary to be at $r = r_0 = 1.014~R_\odot$ (coronal base)
and set the outer boundary at $r=30~R_\odot$ to ensure that the outflow is supersonic.

At the inner boundary $r=r_0$, 
the mass density $\rho$, temperature $T$, and radial magnetic field $B_r$ are fixed to the following values
throughout the computation:% Paramters similar to frequency dependent onset ... by Shoda, Yokoyama, Suzuki
\begin{align}
    \rho_0 = 8.5\times 10^{-13}~\mathrm{kg}~\mathrm{m}^{-3}, ~T_0 = 8\times 10^5~\mathrm{K}, ~B_0 = 10 \, \mathrm{G.}
\end{align}
The radial flow speed is allowed to adjust dynamically and is implemented by enforcing a constant mass flux close to the boundary. At the outer radial boundary, an outflow condition is employed by linearly extrapolating all dynamic quantities. This approach is valid as long as the flow is super-magnetosonic. 

\Alfven{A} waves are introduced into the low corona by utilising time-dependent boundary conditions at $r=r_0$. This is accomplished by specifying the 
Els\"{a}sser $\mathbf{z}^\pm_\perp$ variables transverse to the flux tube, defined by
\begin{align}
        \mathbf{z}^\pm_\perp = \mathbf{v}_\perp \mp \frac{\mathbf{B}_\perp}{\sqrt{\mu_0 \rho}},
\end{align}
at the lower boundary. With this sign convention, $\mathbf{z}^+$
($\mathbf{z}^-$) refer to outgoing (incoming) \Alfven{a} waves propagating along an outward-directed mean field. 
Previous works~\citep{suzuki2004coronal, hollweg1982heating} have noted that linearly polarised \Alfven{a} waves dissipate by direct steepening to MHD fast shocks in addition to turbulent dissipation as a consequence of exciting the PDI. Thus, we study the response of the corona on injecting single-frequency, circularly polarised pump waves, which are direct solutions of the MHD equations~\citep{goldstein1978instability} and dissipate primarily by the PDI, thereby making dissipation less efficient~\citep{suzuki2006solar}. Thus, at the corona base, we specify 
the outgoing Els\"{a}sser variable as
follows:
\begin{align}
        \mathbf{z}^+ = A\sin{\left(2\pi f_0 t\right)}~\mathbf{e}_x + A\cos{\left(2\pi f_0 t\right)}~\mathbf{e}_y.
\label{eq:pump-wave}
\end{align}
Here, $A$ is the amplitude of the perturbations, while $\mathbf{e}_x$ and $\mathbf{e}_y$ denote the unit vectors transverse to the direction of the magnetic field along the flux tube. In this study, we perform a parametric study by considering  injected pump waves with frequencies $f_0 = 0.3, 0.5, 0.66, 1, 2, 3,~$ and $4~$ mHz. The amplitude of the waves are chosen as in \cite{shoda2018frequency} so that $A= 2 \langle v \rangle$ where $\langle v \rangle = 32 \, \mathrm{km/s}$ is the RMS velocity fluctuation amplitude. Finally, the inward waves $\mathbf{z}^-$ are set equal to zero at the inner boundary.

%%%%%%%%%%%%%%%%%%%%%%%%%%%%%%%%%%%%%%%%%%%%%%%%%%%%%%%%%%%%%%%%%%%%%%%%%%%

The calculations are performed in two steps. First, starting from an initial condition representing a hydrostatic equilibrium with a linear flow speed, Equations~\ref{eg:density} to~\ref{eg:energy} are integrated in time without \Alfven{A} wave injections until a steady-state solution is obtained. The steady state solution is then used as the initial condition for a second calculation in which a pump wave is injected.

\subsection{Numerical methods}

The time-dependent MHD equations are solved using a spatially and temporally second-order accurate method employing the Harten–Lax–van Leer (HLL) approximate Riemann solver supplied by piece-wise, linear slope-limited interface states. The semi-discretised equations are advanced in time using the strong stability preserving (SSP) Runge-Kutta method. The same robust methods have been applied in previous studies of the solar corona~\citep{pomoell2012influence}.

The solar wind profile obtained from the simulation is additionally dependent on the number of cells that are used to discretise the simulation domain. We performed multiple simulations, with the number of cells varying from 500 to 5000, finding that the solar wind response remains practically unchanged when more than 3000 cells are used. Thus, the simulation domain is discretised in this study, using 3000 cells spaced logarithmically from the lower corona.

%%%%%%%%%%%%%%%%%%%%%%%%%%%%%%%%%%%%%%%%%%%%%%%%%%%%%%%%%%%%%%%%%%%%%%%%%%%%
\section{Results}    \label{sec:results}
%%%%%%%%%%%%%%%%%%%%%%%%%%%%%%%%%%%%%%%%%%%%%%%%%%%%%%%%%%%%%%%%%%%%%%%%%%% 

\begin{figure*}[h!]
\centering
\includegraphics[width=0.8\textwidth]{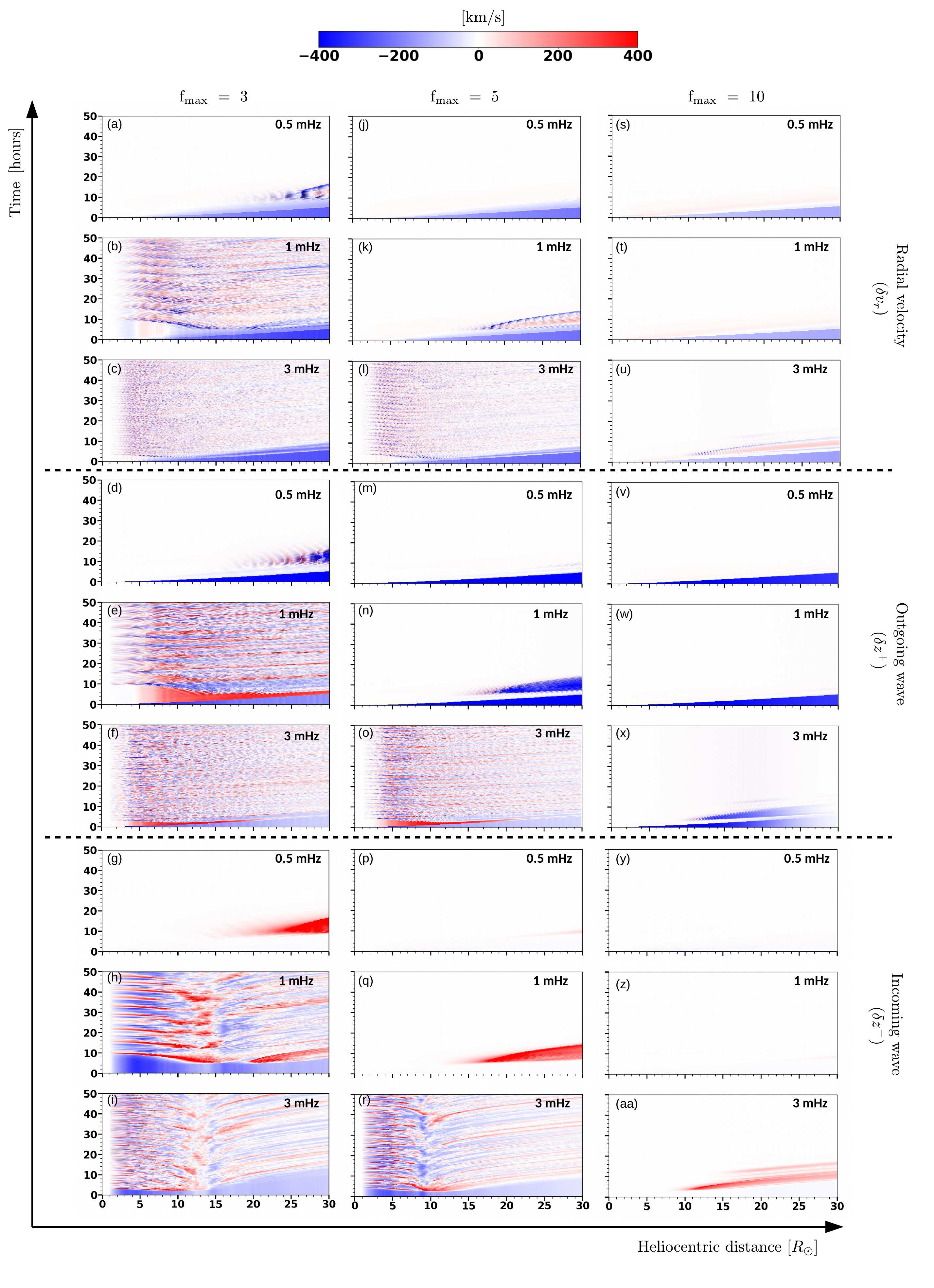}
\caption{Evolution of perturbations in radial velocity $\delta v_r$ (top three rows), magnitude of the 
outgoing Els\"{a}sser variable $\delta z^+$ (middle three rows),
and magnitude of the incoming (towards the Sun) Els\"{a}sser variable $\delta z^-$ (bottom three rows). 
For each quantity, the results for the flux tube expansions, $f_\mathrm{max} = 3, 5, and 10,$ are shown in the left, middle, and right columns, respectively. 
The frequency of the injected pump wave is indicated in the inset of each panel. 
%
%and the magnitudes of the outgoing and incoming Elsässer variables ($\delta z^\pm$) for the various magnetic flux tube expansions. 
%The plots present the evolution of these perturbations in simulation time at the different heliocentric distances for the pump wave frequencies $0.5,~1,~3$~mHz.
}
\label{fig:simulation}
\end{figure*}

\subsection{Steady-state solar wind}    \label{subsec:stead-state-wind}

In Figure~\ref{fig:init-cond}, the state of the corona is shown both before the introduction of \Alfven{a}ic perturbations (panels a-c),  
%state of the simulation before we inject a pump frequency, 
as well as after the solar wind has reached a quasi-steady state upon the injection of \Alfven{a} waves of the lowest considered pump frequency of 0.3 mHz (panels d-f). The solar wind density, sound speed, and radial speed are shown at these two evolutionary instances for the three flux-tube geometries considered, with expansions of $f_\mathrm{max} = 3,~5,~$ and $~10$. The response of a polytropic solar wind ($\gamma < 5/3$) for different magnetic flux tube expansions in the absence of a pump wave has been described in previous works~\citep{kopp1976dynamics, pinto2016flux}. 
%, keppens1999numerical}. 
Close to the Sun ($r < {\sim} 2 R_S$), the corona is seen to exhibit similar plasma conditions irrespectively of the chosen flux-tube geometry. The heliocentric distance to which these similar plasma conditions persist ($r < {\sim} 2 R_S$ in our study) is partly due to the choice of parameters in Equation~\ref{eq:f}. At greater heliocentric distances, the mass density is higher for smaller values of $f_\mathrm{max}$, while the opposite is observed for the sound speed and radial velocity. Upon injecting a $0.3~$mHz \Alfven{a} wave, we see that the solar wind exhibits no visually discernible perturbations, indicating that this medium-frequency (0.3~mHz) \Alfven{a} wave simply acts as a direct source of momentum (via the wave pressure gradient) and energy to the solar wind. 
%In such a scenario the pump wave contributes a constant wave pressure that heats the solar wind. 
As compared to the steady state, the sound speed and density are lower and radial speed higher for all considered $f_\mathrm{max}$. This behaviour of the solar wind in response to hour-scale \Alfven{a} waves is similarly found in other numerical simulations~\citep{shoda2018frequency} and is indicative of the low dissipation of \Alfven{a} waves at these frequencies either by reflections in the medium or by PDI. Such hour-scale \Alfven{a} waves are also seen to dominate observations~\citep{belcher1971large} as these waves do not dissipate and propagate through the medium as in Figure~\ref{fig:init-cond}. The acceleration of the solar wind is discussed in Section \ref{subsec:fmax-scaling}.

\subsection{High-frequency injection: Transient and persisting perturbations}    
\label{subsec:high-freq-inj}

When the frequency of the injected pump wave is increased,
%higher frequencies in turn the injected pump wave starts
persistent large-amplitude perturbations in the solar wind solution are observed. In Figure~\ref{fig:simulation}, we show the 
%propagation of perturbations 
deviation of the
radial velocity and the outward and inward Els\"{a}sser variables from their time-averaged mean values. Specifically, a quantity $q$ (representing either $v_r$, $z^+=|\mathbf{z}^+|$ or $z^-=|\mathbf{z}^-|$) is decomposed according to
\begin{align}
        q = q_0 + \delta q,
\label{eq:averaging}
\end{align}
where $q_0$ is the mean value and $\delta q$ represents the deviation from the mean, that is, the perturbation. The mean value $q_0$ is calculated by averaging in time from the when the simulation likely reaches a quasi-steady state until the end of the simulation. Thus, we average starting from $\tau_0 = 24$ to $\tau_1 = 50~$hours. The figure presents the progression of time 
%is presented as a collection of heatmaps where the 
of the perturbations, with the horizontal axis denoting heliocentric distance and the vertical axis the simulation time elapsed (in hours). For each quantity ($v_r,~z^+,~z^-$), we display a separate 
plot for the three flux tube geometries considered, %$f_\mathrm{max}$ 
as well as for three selected pump frequencies of $0.5,~1, and~3$~mHz. Finally, each of these perturbations are represented in the $[-400,~400]~$km~s$^{-1}$range. 

The perturbations induced by the injected monochromatic pump waves require about $10~$hours to propagate across the simulation domain. 
%following which we can comment on the response of the solar wind. 
This is observed by the initial appearance of a blue region in $\delta z^+$ and $\delta v_r$ for $r > 2 R_\odot,$ signaling a depletion in the quantities. This change is subsequently observed to  disappear at $t \sim 10$~hours. After this initial reconfiguration phase of the corona as a response to the injected waves, perturbations start to appear with a clear dependence both on the pump frequency and flux tube geometry $f_\mathrm{max}$. Perturbations become more persistent with increasing frequency; the frequency at which perturbations become persistent increases with $f_\mathrm{max}$. For instance,  $\delta v_r,$ for the lowest investigated pump frequency $0.5$ mHz in Figure~\ref{fig:simulation} (panels a, j, s), exhibits perturbations only in the case of $f_\mathrm{max} = 3$ from $10-20$~hours. These perturbations for $f_\mathrm{max} = 3$ are advected outwards and cease to be excited, after which point the solar wind attains a steady state. However, these transient perturbations do not arise for $f_\mathrm{max} = 5,~10$. The same behaviour 
%of $\delta v_r$ 
is also detected for $\delta z^\pm$ at $0.5~$mHz. In contrast, for the pump frequency $1~$mHz we observe perturbations that persist for the entire duration of the simulation once excited for $f_\mathrm{max} = 3$, 
%throughout the parameter space, 
while only transient 
perturbations exist for $f_\mathrm{max} = 5$, similar to the case of $0.5~$mHz with $f_\mathrm{max} = 3$ discussed above. Subsequently, with 
an additional increase in the 
%another step up in 
frequency of the pump wave to $3~$mHz, we observe that transient perturbations are generated for $f_\mathrm{max} = 10$, but persist for both $f_\mathrm{max} = 3$ and 5.

\begin{figure*}
\centering
\includegraphics[width=\textwidth]{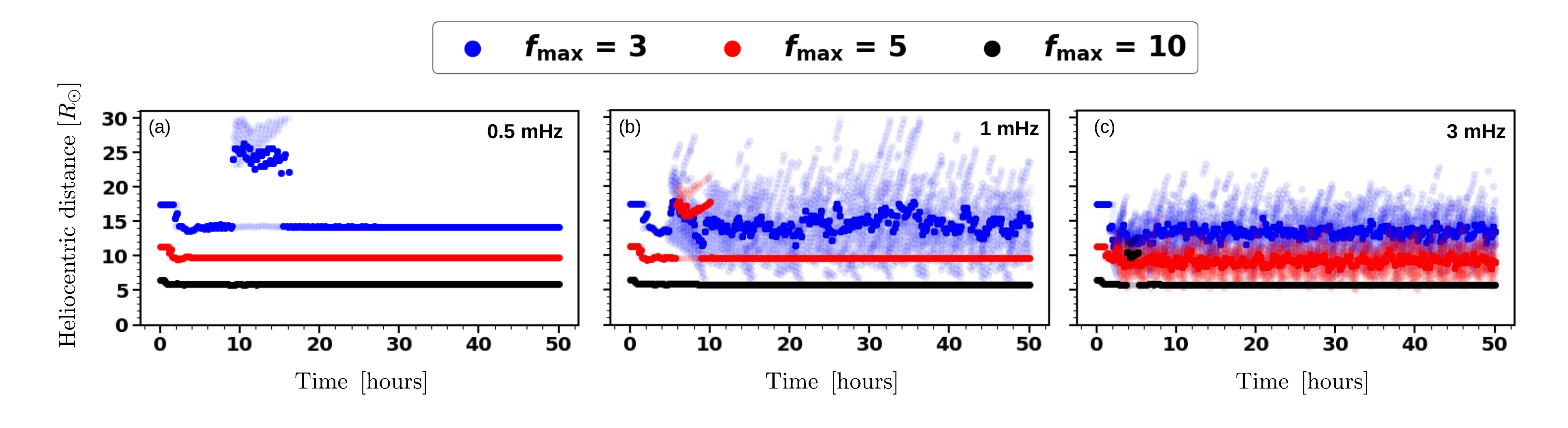}
\caption{%The figure presents 
Evolution of \Alfven{a} crossings %($v_r > v_{\Alfven{a}}$) 
($v_r > v_{A}$) 
with time for the solar-wind conditions in Figure~\ref{fig:simulation}. The opaque markings denote the mean location of these crossings at a given heliocentric distance, while the transparent markings denote the instantaneous locations of the \Alfven{a} crossings.}
\label{fig:crossings}
\end{figure*}

The response of the solar wind for a given flux tube expansion factor as it transitions from transient to persistent perturbations can be better understood through Figure~\ref{fig:crossings}. Here, the locations where the radial velocity equals the local \Alfven{a} speed 
%($v_r > v_{\Alfven{a}}$) 
($v_r = v_{A}$)
are plotted in time for the various flux tube geometries. The absolute locations of the crossings are denoted by partially transparent markings, while the arithmetic mean of the locations at a given time are plotted using opaque markings. In Figure~\ref{fig:simulation}, we observe the solar wind exhibiting no large-amplitude perturbations for a 0.5~mHz pump wave and when $f_\mathrm{max} \geq  5$. Correspondingly, the \Alfven{a} point remains constant after $10~$hours as seen in Figure~\ref{fig:crossings} (panel a), except where we have transient perturbations. Thus, in the absence of large-amplitude solar wind perturbations the injection of a pump wave causes the \Alfven{a} point to occur at a lower heliocentric distance, with this decrease being more prominent for a lower $f_\mathrm{max}$. This greater acceleration of the solar wind for a lower flux-tube parameter is similarly observed in Figure~\ref{fig:init-cond}. Upon injecting a 0.5~mHz pump wave, the \Alfven{a} point moves from $\sim 17.5~R_\odot$ to $\sim 14.5~R_\odot$ for $f_\mathrm{max} = 3$, and from $\sim 6~R_\odot$ to $\sim 5~R_\odot$ for $f_\mathrm{max} = 10$ (Figure~\ref{fig:crossings}). A similar behaviour is also observed for the case of $f_\mathrm{max} = 10$ at 1~mHz. In the case of the short-lived transient perturbations observed from $10-20$~hours for the cases of $f_\mathrm{max} = 3$ at 0.5~mHz, $f_\mathrm{max} = 5$ at 1~mHz, and $f_\mathrm{max} = 10$ at 3~mHz, we see a sharp deviation in the \Alfven{a} point from the initial steady-state wind. The presence of large-amplitude sustained perturbations in the solar wind causes the \Alfven{a} point to oscillate around the unperturbed locations. For instance, in the case of $f_\mathrm{max} = 3$ at 1~mHz, the \Alfven{a} point oscillates approximately around the same location as in the case of $f_\mathrm{max} = 3$ at 0.5~mHz. Further comparing the evolution of the \Alfven{a} points for $f_\mathrm{max} = 3$ and $f_\mathrm{max}=5$ at $0.5~$mHz with the highest $3$~mHz frequency, we observe that the oscillation amplitudes of the \Alfven{a} point decrease with an increase in $f_\mathrm{max}$ and pump frequency. 

The analysis of \Alfven{a} crossings reveals that short-lived transient perturbations for $\delta v_r$ and $\delta z^\pm$ in Figure~\ref{fig:simulation} occur as the solar wind advects any reflected waves that are excited beyond the \Alfven{a} point outwards. In Figure~\ref{fig:simulation}, we see that incoming waves ($z^-$) are generated beyond $\sim 15R_\odot$ at $t = 8~$hours for a pump frequency of 0.5~mHz and $f_\mathrm{max} = 3$. The corresponding \Alfven{a} point at this time is $\sim 14R_\odot$ (Figure~\ref{fig:crossings}a). As a consequence, this disturbance is advected outwards by the solar wind, resulting in the short-lived transient perturbations and the jump in the \Alfven{a} point to $\sim 25~R_\odot$ between $10~and 20~$hours. This reasoning similarly applies for $f_\mathrm{max} = 5$ at $1~$mHz, and $f_\mathrm{max} = 10$ at $3~$mHz.

\subsection{Reflected \Alfven{a} waves in the solar wind}
\label{subsec:reflected-AW}
In Section~\ref{subsec:high-freq-inj}, 
we observed the response of the solar wind upon the injection of a pump wave and the generation of corresponding perturbations leading to varying \Alfven{a} points. Here, we investigate the generation mechanism of the inward propagating ($\mathbf{z^-}$) \Alfven{a} wave that is formed as a response to injecting the outward pump wave at certain frequencies. In Figure~\ref{fig:simulation}, we observed the presence of corresponding perturbations in $z^-$ , which correlated with persisting perturbations in $z^+$ and $v_r$. As discussed in 
Section~\ref{sec:introduction}, \Alfven{a} 
waves can be reflected due to both the inhomogeneity of the medium as well as the PDI.

In order to quantify the presence of \Alfven{a} waves, we consider the time-averaged component of the Els\"{a}sser variables (i.e. $z_0^\pm$). The presence of increased \Alfven{a} wave reflections creates suitable conditions for turbulent heating and slow mode generation by modulating the magnetic field, which can lead to fast shocks~\citep{goldreich1995toward, uchida1974excess, wentzel1974coronal, cranmer2015driving, suzuki2004coronal}. These processes in turn contribute to the further increase in density perturbations. As a consequence, it is possible to generate density perturbations in the solar wind, even in the absence of the PDI. To capture the presence of a compressive wave mode we define the time-averaged fractional density fluctuation $n$ as~\citep{shoda2018self}
\begin{align}
        n = \frac{1}{\langle\rho\rangle}\sqrt{\langle\left(\rho^2 - \langle\rho\rangle\right)^2\rangle},
        \label{eq:n}
\end{align}
where $\langle x\rangle$ denotes the time-averaged value of $x$ computed as $\langle x\rangle = 1/(\tau_1 - \tau_0)\int_{\tau_0}^{\tau_1} x~dt$. Here, we set $\tau_0 = 24$ hours and $\tau_1 = 50~$hours. We note that this averaging procedure was performed differently from that used in Equation~\ref{eq:averaging} as the averaging is here computed dynamically during the simulation run-time using each time-step as an input date. Through defining $n$ as such, we are able to capture the perturbations around the averaged density $\langle\rho\rangle$. According to~\citet{cranmer2012proton}, the fractional density fluctuations without PDI and only with reflections is given by $\delta \rho_\mathrm{rms}/\rho \lessapprox 0.1$. Thus, we take the threshold for the large-scale density perturbations arising due to PDI to be of the order of $n > 0.1$. Subsequently, the presence of any non-zero $z_0^-$ in the absence of large-scale density perturbations indicates \Alfven{a} wave reflections due to medium inhomogeneity.  

\begin{figure*}[h]
\centering
\includegraphics[width=\textwidth]{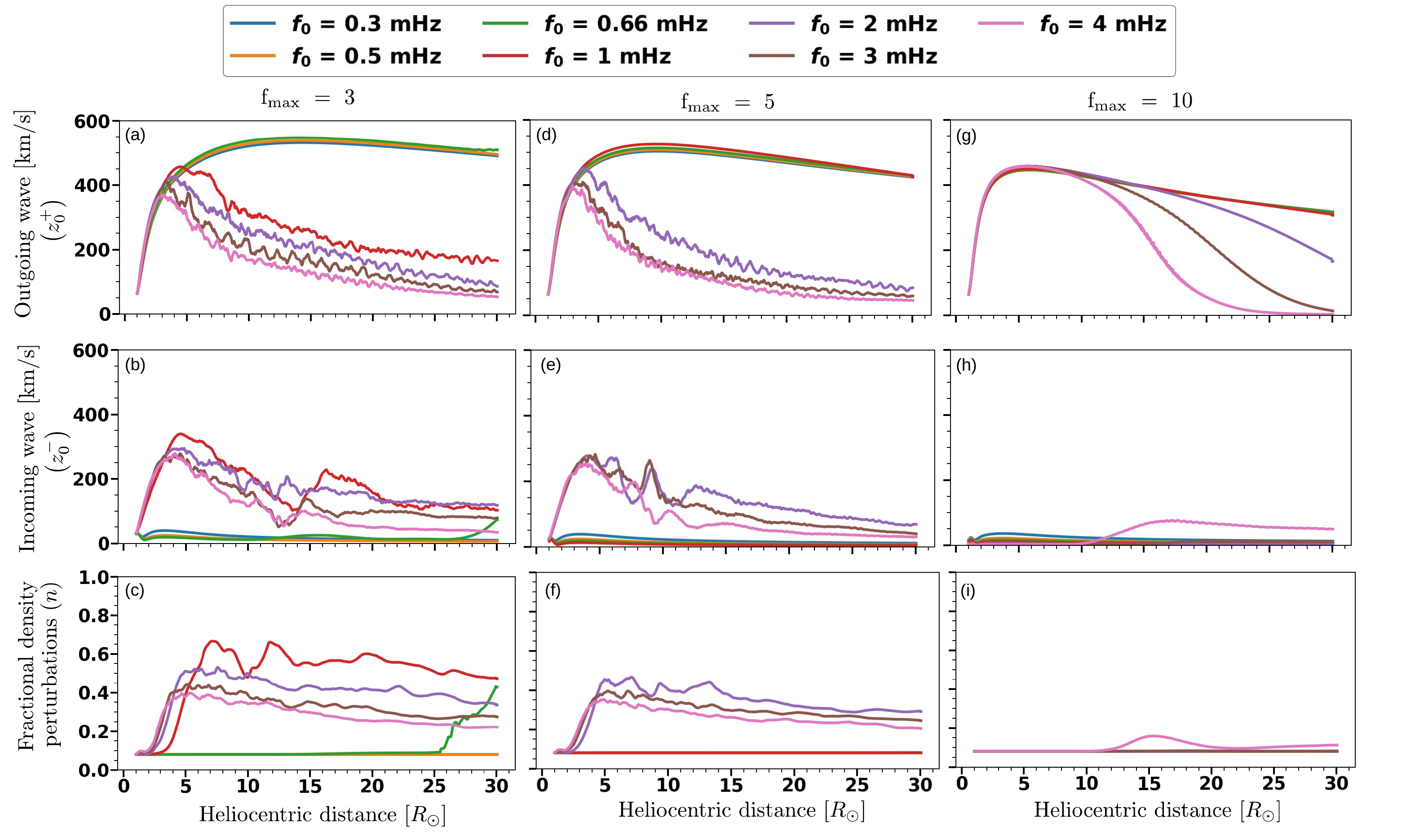}
\caption{Time-averaged outgoing and incoming Els\"{a}sser variables ($z_0^\pm$) and the fractional density fluctuation ($n$) (bottom row) for all the pump-wave frequencies and magnetic flux-tube expansions considered. }
\label{fig:pdi}
\end{figure*}

In Figure~\ref{fig:pdi}, we show the time-averaged outgoing and incoming Els\"{a}sser variable magnitudes and the fractional density fluctuation $n$ for all the considered values of $f_\mathrm{max}$ and pump frequency. At low frequencies of $f = 0.3$ mHz, there is minimal development ($<~50~$km~s$^{-1}$) of the incoming $z^-_0$ for all the %$f_\mathrm{max}$ 
%values 
flux tubes.
%and the solar wind is accelerated by the geometry of the magnetic flux tube~\cite{kopp1976dynamics}, 
%and the constant wave pressure generated by the pump wave. 
As the frequency of the pump wave increases, the development of $z^-_0$ is clearly observed (panels b, e, and h). However, the threshold frequency at which 
%the incoming els\"{a}sser variable begins 
$z^-_0$ reaches larger amplitudes
%to be excited 
varies with $f_\mathrm{max}$. For instance, it is $\approx 0.66~$mHz for $f_\mathrm{max} = 3$ and $\approx 4~$mHz for $f_\mathrm{max} = 10$. In addition, for a given $f_\mathrm{max}$ value, we observe significant density perturbations for the frequency ranges that exhibit  larger peak values for $z^-_0$. These corresponding features in $z_0^-$ and $n$ therefore indicate the development of the PDI. For instance, the instability is excited after $\sim 25~R_\odot$ for $f_\mathrm{max} = 3$ at $0.66~$mHz and after $\sim 13~R_\odot$ for $f_\mathrm{max} = 10$ at $4~$mHz. %Alternatively, we observe the presence of some $\mathbf{z_0^-}$ for $f = 0.5~$mHz and $f_\mathrm{max} = 3$ without the presence of density perturbations indicating that these reflections are generated by the medium inhomogeneity's. 

The mechanism through which reflected \Alfven{a} waves are generated, and particularly the results presented in Figures~\ref{fig:simulation}~\&~\ref{fig:crossings}, may be identified with PDI. As the quantities shown in Figure~\ref{fig:pdi} are averaged from 24~hours to 50~hours, they do not show %are unable to contain 
the transient perturbations that occur in the simulation before this time. However, a similar analysis 
%of our results 
indicates that short-lived, large-amplitude density perturbations are associated with these phenomena, showing that the transient perturbations are caused by the excitation of the PDI and with the resulting waves advected outwards by the solar wind, since the location at which the instability is triggered is beyond the \Alfven{a} point.
%it is the PDI location that is advected outwards by the solar wind in such cases. 
In contrast, the presence of persistent perturbations in Figure~\ref{fig:simulation} are generated by the PDI occurring below the  \Alfven{a} point, thereby allowing the waves to remain in the subsonic solar wind. For instance, for the case of $f_\mathrm{max} = 5$ at 3~mHz, we observe sustained perturbations in Figure~\ref{fig:simulation}, with the corresponding \Alfven{a} point oscillating around $\sim 7 R_\odot$ (Figure~\ref{fig:crossings}c) and the PDI being excited after $\sim 3 R_\odot$.

%system is complicated after the linear to non-linear evolution of the PDI say after 20 hours when it is first excited

\section{Discussion}
\subsection{Onset of the PDI} \label{subsec:onset-pdi}

\begin{figure}
\centering
\includegraphics[width=0.5\textwidth]{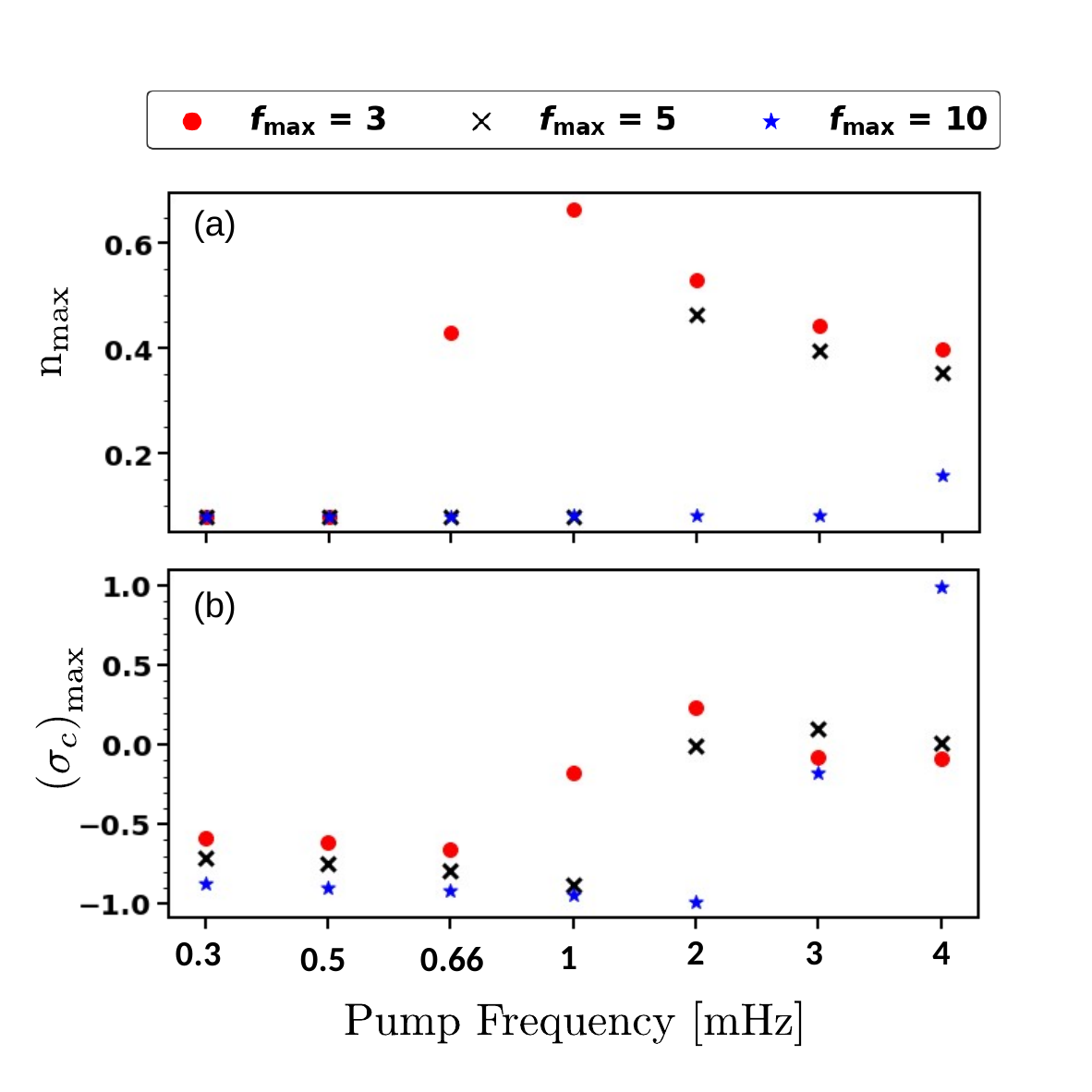}
\caption{Maximum fractional density perturbations ($n_\mathrm{max}$) and cross-helicity ($(\sigma_c)_\mathrm{max}$) for the different pump-wave injection frequencies and flux tube geometries.}
\label{fig:threshold}
\end{figure}

In order to quantify the threshold for the onset of the PDI for the solar wind solutions with different flux tube geometries, we calculate the maximum of the fractional density perturbations $n_\mathrm{max}$ and the maximum of the cross-helicity $(\sigma_c)_\mathrm{max}$. Here, the cross-helicity $\sigma_c$ is defined as
\begin{align}
        \sigma_c = \frac{\langle E^-\rangle - \langle E^+\rangle}{\langle E^-\rangle + \langle E^+\rangle}.
        %~ E^\pm = \frac{1}{2}\rho {z^\pm}^2
\end{align}
%
%with $E^\pm = \frac{1}{2}\rho {z^\pm}^2$ the energy density of the \Alfven{A} waves.
Similarly to the definition of $n$ (Equation~\ref{eq:n}), $\langle E^\pm \rangle$ denotes the time-averaged value of the energy density of the \Alfven{A} waves $E^\pm = \frac{1}{2}\rho {z^\pm}^2$ averaged over one day for all heliocentric distances in the quasi-steady solution. In a situation with no incoming $z^-$ waves, a value of $\sigma_c = -1$ would be obtained. Any departure from this value indicates the presence of a reflected \Alfven{a} wave in the simulation. By comparing the values $n_\mathrm{max}$ and $(\sigma_c)_\mathrm{max}$ for the range of pump frequencies and flux tube areas considered, we can investigate the dependence of the threshold for the excitation of the PDI on $f_\mathrm{max}$ (Figure~\ref{fig:threshold}). These parameters are indicators of the PDI as this instability generates large-amplitude density fluctuations, which increase $n$, and reflected \Alfven{a} waves, which increase $\sigma_c$.

In Figure~\ref{fig:threshold}, we observe the dependency of the onset of the PDI on the flux tube geometry. Previous works~\citep{Ferraro1958, an1990reflection, velli1993propagation, cranmer2005generation, hollweg2007reflection} have noted that the reflection of \Alfven{a} waves due to the medium itself is more efficient at lower frequencies. This behaviour is also %observed in %this study 
consistent with the behaviour observed in
Figure~\ref{fig:threshold},
where the presence of incoming ($z^-$) waves for 0.3 and 0.5~mHz is indicated by the non-zero values of $(\sigma_c)_\mathrm{max,}$ while simultaneously no large-scale density perturbations are present for all the $f_\mathrm{max}$ values. For smaller frequencies of the pump wave,
$(\sigma_c)_\mathrm{max}$ is larger, and it progressively decreases until a pump frequency of approximately $0.66~$mHz is reached. Thus, we observe that increasing the pump frequency causes a reduction in the generation of reflected \Alfven{a} waves, with the trend continuing  monotonically up to a threshold frequency, with the threshold depending on the flux-tube geometry. For instance, a clear jump in $n_\mathrm{max}$ and $(\sigma_c)_\mathrm{max}$ is observed to occur for $f_\mathrm{max} = 3$ at $0.66~$mHz. This breakdown from the monotonic trend for $f_\mathrm{max} = 5,~10$ is observed at $2~$mHz and $4~$mHz, respectively. Thus, Figure~\ref{fig:threshold} indicates the thresholds for the onset of the PDI as the point at which the monotonically decreasing trend of $(\sigma_c)_\mathrm{max}$ breaks down with an accompanying increase in $n_\mathrm{max}$. 

\begin{figure*}
\centering
\includegraphics[width=\textwidth]{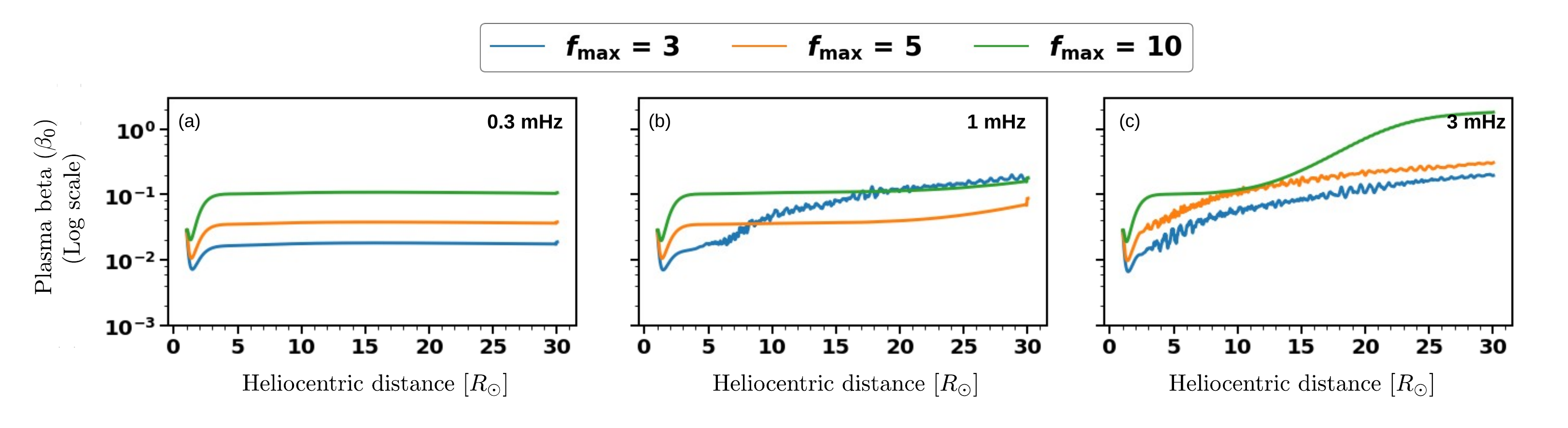}
\caption{Time-averaged plasma beta ($\beta_0$) for (a) 0.3~mHz, (b) 1~mHz, and (c) 3~mHz are presented.}
\label{fig:plasma-beta}
\end{figure*}

This suppression of PDI by the flux-tube expansion factor ($f_\mathrm{max}$) can be explained by considering the plasma beta (Figure~\ref{fig:plasma-beta}). Here, the plasma beta ($\beta$) is time-averaged, as defined in Equation~\ref{eq:averaging} and plotted for pump frequencies $0.3,~1, and~3~$mHz. The plasma beta is defined as $\beta = P/(B^2/2\mu_0)$, where $P$ is the total thermal pressure and $B$ is the magnitude of the total magnetic field. In the absence of PDI, we observe the  $\beta \ll 1$ conditions as the solar wind expands (Figure ~\ref{fig:plasma-beta}(a)). The subsequent large-amplitude fluctuations generated by the PDI instability increase the plasma beta as seen for $f_\mathrm{max} = 3$ in Figure~\ref{fig:plasma-beta}(b) and $f_\mathrm{max} = 3,~5,~10$ in Figure~\ref{fig:plasma-beta}(c). Upon comparing Figures~\ref{fig:plasma-beta}(a)~\&~\ref{fig:plasma-beta}(c), we see that plasma beta increases from $<0.1$ to $~2$ for $f_\mathrm{max} = 10$, but only from $\ll 0.1$ to $\approx 0.2$ for $f_\mathrm{max} = 3$. Thus, the plasma beta is small when $f_\mathrm{max}$ is lower, and the relative increase in $\beta$ is smaller for a lower $f_\mathrm{max}$ when the PDI is excited. Previous analytical studies~\citep{galeev1963stability, sagdeev1969nonlinear} have shown the inverse dependence of the PDI growth rate on plasma beta ($\gamma_\mathrm{PDI} \propto \beta^{-1/4}$) in the limit $\beta \ll 1$. Considering the decrease in $\beta$ as the solar wind expands more slowly (smaller $f_\mathrm{max}$), the onset of PDI is suppressed as the magnetic flux tube expansion factor increases. This variation in the PDI growth rate with $f_\mathrm{max}$ is shown by~\citet{tenerani2013parametric}.  
%The suppression of the PDI by the expansion of the solar wind ($f_\mathrm{max}$) and the inhomogenity of the medium has also been discussed by \cite{tenerani2013parametric}.

\subsection{Acceleration of the solar wind} \label{subsec:fmax-scaling}

Comparing the steady-state solar wind solution before the injection of \Alfven{A} waves with 
the quasi-steady-state solution obtained upon the injection of low-frequency waves, a clear difference in the resulting solar wind speed is seen in Figure~\ref{fig:init-cond}. In particular, the terminal wind speed is found to increase with increasing expansion $f_\mathrm{max}$. The same trend is also observed when injecting low-frequency perturbations (Figure~\ref{fig:init-cond}f), although the difference in the terminal speeds is smaller in this case. 

%Figure~\ref{fig:init-cond} 
The wind speeds corresponding to the different flux-tube expansion factors appear contradictory to 
observations and empirical modelling in which the wind speed has been found to be inversely correlated with the expansion factor~\citep{wang1990solar}. 
%In reality the acceleration of the solar wind for a lower $f_\mathrm{max}$ parameter is higher. 
This apparent difference is caused by the fact that we do not explicitly account for turbulent heating in our present numerical model. In our study, the solar wind is primarily accelerated due to the wave pressure generated by the injected pump wave and the reflected waves, and compressive heating. For a given frequency and flux tube geometry ($f_\mathrm{max}$ value), the pump wave is continually injected at a constant amplitude and frequency. This injected pump wave further excites alternate wave modes as a consequence of the PDI (for sufficiently large frequencies). Additionally, we note that a circularly polarised pump wave results in minimal dissipation of heat through direct steepening of outgoing waves to fast shocks~\citep{suzuki2004coronal, hollweg1982heating}. Thus, the variance in the solar-wind acceleration and heating is dependent on the extent of reflected waves. In Figure~\ref{fig:init-cond}, we observe that an injection of a 0.3~mHz pump wave causes the wind speed for $f_\mathrm{max} = 3$ to increase the most due to the presence of higher amount of incoming $\mathbf{z^-}$ waves (see Figures~\ref{fig:pdi},~\ref{fig:threshold}). 
%for this value of $f_\mathrm{max}$ at 0.3~mHz.  
Additionally, we observed that the onset threshold for the PDI occurs at a lower frequency for a smaller $f_\mathrm{max}$ parameter (Figure~\ref{fig:threshold}).
% can the relationship between greater fmax=3 and z- be seen in the momentum equation directly
\begin{figure*}[h]
\centering
\includegraphics[width=\textwidth]{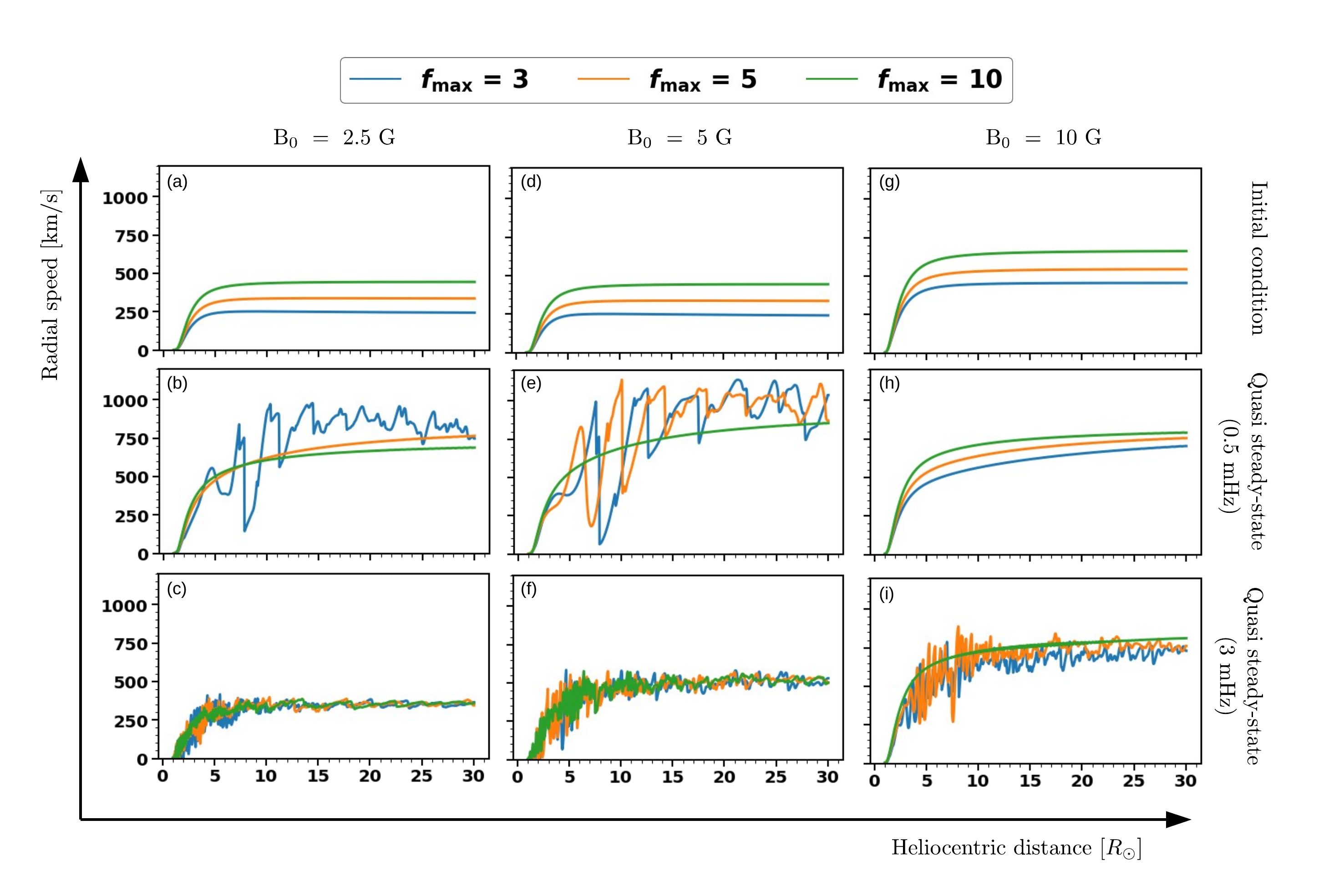}
\caption{Response of the solar wind for different background magnetic fields ($B_0 = 2.5,~5,~10$~G) and pump-wave injection frequencies ($0.5,~3$~mHz) are shown.}
\label{fig:vr_B}
\end{figure*}

In addition to the flux-tube expansion geometry, the solar-wind speed is dependent on the magnetic-field magnitude ($B_0$), as has been noted, or example,i by \cite{suzuki2004coronal}. In Figure~\ref{fig:vr_B}, we compare the radial speed for the various flux-tube geometries for different field strengths varied according to $B_0 = 2.5,~5,~$and$~10~$G. For the case when a pump wave is not injected, we see that $f_\mathrm{max} = 10$ consistently attains the highest terminal speed. Upon the injection of a 0.5~mHz pump wave, we observe the presence of significant perturbations in the radial speed for $f_\mathrm{max} = 3$ at $B_0 = 2.5~$G and $f_\mathrm{max} = 3,~5$ at $B_0 = 5~$G. Similarly to the development of perturbations in Figure~\ref{fig:simulation}, these perturbations of the radial speed correspond to the generation of reflected waves and density perturbations, indicating the development of the PDI for the two cases outlined above. Thus, we observe the presence of greater amounts of reflected waves for a lower $B_0$ as seen in the magnetic field strength (Figure~\ref{fig:vr_B}) at which we start observing perturbations in $v_r$ for a given $f_\mathrm{max}$ parameter. It is to be noted that the case of $f_\mathrm{max} = 5$ at $2.5~$G is similar to the short-lived transient perturbations observed in Figure~\ref{fig:simulation}, where the instability occurs after the \Alfven{A} point. Additionally, the injection of the $0.5~$mHz pump wave decreases the relative differences in the speeds for different $f_\mathrm{max}$ parameters, even in the case when the PDI is not excited ($0.5~$mHz pump wave at $f_\mathrm{max} = 10$). As stated previously, this variation in speeds is a consequence of the extent of reflected waves generated in our simulation. Subsequently, additional momentum is supplied to the solar wind through the development of alternate wave modes through the development of reflected $z^-$ waves, and compressive wave modes. Therefore, for a $0.5~$mHz injection the relative scaling of radial speeds for different flux-tube geometries remains the same for $B_0 = 10~G,$ while for $B_0 = 2.5,~5~$G the greatest $v_r$ is observed for $f_\mathrm{max} = 3, 5$. Finally, in the case with a $3~$mHz pump wave that causes the PDI to be excited for all the considered flux tube expansions, we see that the relative difference between the radial speeds for a given field strength $B_0$ shrinks further. Therefore, if we explicitly included a model for turbulent heating in our model the radial speed for $f_\mathrm{max} = 3$ would be expected to be the largest, as in this case we observe the presence of a greater extent of reflected waves as seen by the relative increase in $v_r$ from the case without injection of waves. Consequently, the injection of pump waves in a broadband spectrum of frequencies would result in a greater presence of reflected waves for a lower $f_\mathrm{max}$ and $B_0$ contributing to a higher acceleration of the solar wind. %This reasoning is consistent with previous simulation works noting the importance of turbulent heating in order to achieve the fast solar wind~\cite{shoda2018self}.

%%%%%%%%%%%%%%%%%%%%%%%%%%%%%%%%%%%%%%%%%%%%%%%%%%%%%%%%%%%%%%%%%%%%%%%%%%%%
\section{Conclusion}    \label{sec:conclusion}
%%%%%%%%%%%%%%%%%%%%%%%%%%%%%%%%%%%%%%%%%%%%%%%%%%%%%%%%%%%%%%%%%%%%%%%%%%%
In this study, we investigated the response of the solar wind upon the injection of a monochromatic pump \Alfven{a} wave. The study is performed considering different magnetic flux-tube expansion geometries (parametrised by $f_\mathrm{max}$) using a single-fluid MHD simulation, which models the non-linear dynamics caused by the pump wave introduced using time-dependent boundary conditions. The model accounts for compressive heating for the range of pump frequencies used in this paper. Thus, given an injection frequency, the model continually injects wave pressure at the low coronal boundary and transfers momentum to the solar wind. This injected \Alfven{a} wave propagates in the solar wind, giving rise to reflected \Alfven{a} waves and compressive wave modes due to the PDI and the inhomogeneity of the medium. Upon injecting the pump wave in quiet solar-wind conditions (Figure~\ref{fig:init-cond}), we observed the development of both short-lived and persistent perturbations in our solar-wind solution. An investigation of the reflected \Alfven{a} waves and large-scale density perturbations in Sections~\ref{subsec:high-freq-inj} \&~\ref{subsec:reflected-AW} indicate the development of the PDI. The solar-wind response and the resulting perturbations (Section~\ref{subsec:high-freq-inj}) were then explained by the relative positions of the \Alfven{a} point (Figure~\ref{fig:crossings}) and the PDI location, which results in the solar wind advecting the disturbances that occur in the super-\Alfven{a}ic region outwards. 

Previous works have investigated the dependence of the onset of the PDI on the pump frequency for a single $f_\mathrm{max}$~\citep{shoda2018frequency, reville2018parametric} and commented on the suppression of the PDI at low frequencies due to efficient reflections of these wave modes. In our study, we found a similar tendency that can be seen in the break of the monotonic trend of $(\sigma_c)_\mathrm{max}$ with the pump frequency in Section~\ref{subsec:onset-pdi}. Furthermore, our study finds a strong relationship between the flux-tube expansion and the resulting \Alfven{a} reflections and solar-wind acceleration. In Section~\ref{subsec:onset-pdi}, we discussed the PDI threshold for different $f_\mathrm{max}$ values which is consistent with the suppression of the PDI by the magnetic flux tube geometry~\citep{tenerani2013parametric}. Additionally, in Section~\ref{subsec:fmax-scaling} we present results that indicate the suppression of the PDI by the increase in magnitude of the background magnetic field (Figure~\ref{fig:vr_B}, panels b, e, and h). However, our results displayed an inconsistency with observational results comparing terminal wind speeds for different flux tube expansion factors~\citep{wang1990solar}. This anomaly is discussed in Section~\ref{subsec:fmax-scaling}, and the difference was noted to be a consequence of not including an explicit model for turbulent heating in our model. This investigation indicates the importance of considering turbulent heating in order to achieve realistic solar-wind speed profiles, as cited in other works~\citep{suzuki2005making, shoda2018self}. 

Based on our results, the $1/f$ scaling of the terminal speed of the solar wind is intimately linked to the turbulent heating, which itself depends intricately on the flux-tube expansion and magnetic field strength. Understanding the coupling of these is a key step in advancing our understanding of the importance of turbulent heating and other processes in accelerating the solar wind. 

In the process of performing our study, we experimented with alternate values of the pump-wave amplitude (Equation~\ref{eq:pump-wave}) and the parametrisation of the magnetic flux-tube expansion via $R_1$ and $\sigma_1$(Equation~\ref{eq:f}). We found that the pump-wave amplitude also inhibits PDI development, with the instability requiring the presence of large amplitude \Alfven{A} waves as considered in our study. Furthermore, the parametrisation in Equation~\ref{eq:f} affects the location at which PDI is excited. The parameters $R_1$ and $\sigma_1$ in Equation~\ref{eq:f} represent the location and width of the region experiencing the most variation in the flux tube. Thus, alternate choices of these parameters resulted in variance in the location at which the instability is excited without affecting the qualitative behaviour of the solar wind at the given pump frequency and $f_\mathrm{max}$ value.

The structure of the solar magnetic flux exhibits a varying expansion factor ($f_\mathrm{max}$) dependent on the heliocentric distance~\citep{griton2021source}. In comparison, simulation studies that seek to understand the dynamics of \Alfven{A} wave propagation in the corona employ a fixed $f_\mathrm{max}$ derived from a PFSS approximation of the background field. Additionally, previous studies~\citep{pinto2016flux} have noted the non-linear dependence of terminal wind speeds on the magnetic flux-tube geometry. Considering the suppression of PDI by the flux-tube geometry and background field studied in this paper, the solar-wind response to \Alfven{A}ic fluctuations originating in the photosphere is thus highly sensitive to the approximated magnetic-field geometry. Therefore, through this study we have sought to develop a greater understanding of the validity and formulation of reflection-driven \Alfven{a}ic turbulent heating in global simulations of the solar wind and corona. These investigations of the dynamics of \Alfven{A}-wave propagation in the corona are relevant to understanding the near-sun observations of recent spacecraft missions, such as results~\citep{shoda2021turbulent} suggesting an \Alfven{A} origin for magnetic switchbacks in PSP data.

%%%%%%%%%%%%%%%%%%%%%%%%%%%%%%%%%%%%%%%%%%%%%%%%%%%%%%%%%%%%%%%%%%%%%%%%%%%%
\begin{acknowledgements} 
The work has been supported by the Finnish Centre of Excellence in Research on Sustainable Space (FORESAIL). This is a project under the Academy of Finland.
\end{acknowledgements}

%%%%%%%%%%%%%%%%%%%%%%%%%%%%%%%%%%%%%%%%%%%%%%%%%%%%%%%%%%%%%%%%%%%%%%%%%%%%
%% references
\bibliographystyle{aa-note} %% aa.bst but adding links and notes to references
%%\raggedright              %% for latex+dvips, not for pdflatex
\bibliography{example}      %% example.bib = bibtex entries copied from ADS

\end{document}